\documentclass[twocolumn,twoside,slac_two]{revtex4}
\usepackage{graphicx}
\usepackage{fancyhdr}
\newcommand{\fermi}{\textit{Fermi}/LAT}
\newcommand\arcdeg{^{\circ}}
\newcommand\arcs{^{\prime\prime}}
\newcommand{\arcm}{^\prime}

\begin{document}

\title{Fermi/LAT Study of the Cygnus Loop Supernova Remnant: Discovery of a Point-like Source and of Spectral Differences in its gamma-ray emission}

%

\author{I. Reichardt\footnote{now at INFN Padova, Italy (\texttt{ignasi.reichardt@pd.infn.it}).}}
\author{R. Terrier}
\affiliation{Astroparticule et Cosmologie (APC), Universit\'e Paris 7 Denis Diderot, 75205 Paris Cedex 13, France}
\author{J. West}
\author{S. Safi-Harb}
\affiliation{Department of Physics and Astronomy, University of Manitoba, Winnipeg, MB R3T 2N2, Canada}
\author{E. de O\~na-Wilhelmi}
\affiliation{Institut de Ci\`encies de l'Espai (IEEC-CSIC), 08193 Bellaterra, Spain}
\author{J. Rico}
\affiliation{Institut de F\'isica d'Altes Energies (IFAE), 08193 Bellaterra, Spain}

\begin{abstract}
The Cygnus Loop is a nearby supernova remnant (SNR) observed across the electromagnetic spectrum. With the analysis of 6 years of Fermi/LAT data we find that, what previous studies had considered a single source, consists of an extended source plus a point-like source south-east of the SNR. The extended gamma-ray emission is well correlated with the thermal X-ray emission of the SNR, and the energy spectrum displays a pronounced maximum at $\sim0.6$\,GeV. However, in a region where the radio emission is strongly and distinctly polarized, the gamma-ray spectrum shows no sign of a break. Therefore, the spatially resolved gamma-ray emission permits the study of different interaction conditions of the SNR and the surrounding medium.
\end{abstract}

\maketitle

\thispagestyle{fancy}


\section{INTRODUCTION}
The Cygnus Loop is the remnant of a core-collapse supernova explosion that occurred about 14000 year ago \citep{graham1998} at a distance of $540^{+100}_{-80}$\,pc \citep{CLHubble}. The Cygnus Loop is among the closest supernova remnants (SNRs) to Earth, which implies that it could act as a local accelerator. Due to its proximity, the Cygnus Loop is seen on the sky with an angular size of about 3 degrees. In general, the blast wave of the SNR is not breaking out of a dense cloud, but running into a wall of atomic gas related to the cavity in which the supernova occurred. The wall slows down the shock, which becomes bright in optical emission lines. The reflected shock propagates through the hot interior, which enhances the X-ray emission in correlation with the optical emission \citep{graham1995,levenson1996}. However, some portions of the shock proceed unimpeded through low-density inter-cloud medium.

X-ray emission from reflection-shocked gas is particularly bright in the east. In contrast, the south of the SNR (the so-called \textit{breakout}) is very dim in X-rays. This is often regarded as caused by the expansion of the blast wave into a low-density medium. However, \citet{uyaniker} found that the polarization of the 2695\,MHz emission was much higher there with respect to the north of the shell. A possible interpretation of this feature is that a second SNR is present in that region, and interacts with the Cygnus Loop.

No compact object is firmly associated with the collapsed progenitor of the Cygnus Loop. A few candidates lie within the breakout, where the \textit{ASCA} survey revealed a point-like source, but it is not firmly established as a neutron star \citep{miyata}. There is yet another compact object with a candidate pulsar wind nebula nearby, revealed by \textit{Suzaku} and \textit{XMM-Newton} observations \citep{cygloopPWN}. No pulsations have been detected from any of these objects. In addition, a very high transverse proper motion of $\sim 1300$\,km~s$^{-1}$ is needed if it is assumed that one of these candidate neutron stars departed from the geometric center of the Cygnus Loop some 14000 years ago. Such a supersonic movement would produce a cometary shape in the X-ray emission that has not been observed so far. However, this could be explained if the neutron star was related to the second SNR suggested by \citet{uyaniker}.

The detection of GeV gamma-ray emission from the Cygnus Loop was published in \citet{cygloop}, who analyzed two years of \fermi\, data comprised between August 2008 and August 2010. In this analysis, the shape of the Cygnus Loop was modeled as a ring, somewhat more extended than the shell seen at other wavelengths. The spectrum is curved (modeled as a log-parabola), and the fit to a one-zone hadronic model returns plausible values for the parameters.

In this work we analyze six years of \fermi\, data using the latest software. The factor 3 increase in statistics with respect to the previous study provides unprecedented sensitivity to study both spatial and spectral features of the gamma-ray emission from the Cygnus Loop.

\section{DATA ANALYSIS}
We analyzed \fermi\, \texttt{Pass 7} Reprocessed data corresponding to the period between August $4^{th}$ 2008 (start of science operations) and September $7^{th}$ 2014. We defined the ROI as a circle of $10\arcdeg$ radius centered at the position (RA, DEC) = ($20^{h}58^{m}11^{s}$, $29\arcdeg23\arcm56\arcs$), J2000, which is $2\arcdeg$ displaced towards negative Galactic latitudes with respect to the catalog position of the Cygnus Loop. This is done in order to be less affected by the diffuse emission from the Galactic plane. Data were processed with the version v9r32p5 of the \textit{ScienceTools}. We selected class 3 events in the energy range between 58.5\,MeV and 300\,GeV, with the recommended quality cuts (including the requirement for the spacecraft to be in normal operation mode, LAT\_CONFIG=1, data to be flagged as good quality, DATA\_QUAL=1, and a cut on the rocking angle of the spacecraft, ABS(ROCK\_ANGLE)$<52\arcdeg$). In addition, we applied a zenith angle cut of $100\arcdeg$ in order to prevent event contamination from the Earth limb. Data were binned in sky coordinates with the \texttt{gtbin} tool, using square bins of $0.125\arcdeg$ side. This tool produces a \textit{counts map}, with the number of events recorded by the detector.

We performed a binned likelihood analysis with a model containing the standard Galactic and extra-galactic diffuse emission models provided in the \textit{ScienceTools}, plus the sources in the 2FGL catalog lying up to $15\arcdeg$ away of the ROI center.
We call the model with the point-like sources plus the Galactic and extra-galactic backgrounds the \textit{null hypothesis}, which has a maximum likelihood $\mathcal{L}_{0}$. Then, we generate alternative models by adding spatial templates and by changing the functions describing spectral shape. By varying the parameters of each models, we compute the corresponding maximum likelihood $\mathcal{L}_{model}$. We choose the best representation of the Cygnus Loop as the model which obtains the highest value of the likelihood ratio LR=$2\log(\mathcal{L}_{0}/\mathcal{L}_{model})$.

For any of the tested models, we can use the tool \texttt{gtmodel} to produce an \textit{expected counts map} given the exposure associated to the data set. For visualization purposes, we produce what we call the \textit{S/N map} by subtracting the expected counts map from the actual counts maps, and then dividing by the square root of the expected counts map.

Complementary to the \fermi\, data analysis, we have re-analyzed the 11\,cm radio emission observed by the 100\,m Effelsberg telescope. \citet{uyaniker} proposed the Cygnus Loop be divided in two regions for the two-SNR interpretation. Based on our re-analysis, which considers the presence of extended Stokes I radio emission in addition to the distinct and intensely polarized radio emission, we have re-defined the regions to be more equal in size and both having a circular shape. We consider that the southwest (SW) feature is the circular region of $1.07\arcdeg$ radius centered at (RA, DEC) = ($20^{h}49^{m}$, $29\arcdeg47\arcm$) as shown in Figure~\ref{fig:CLmulti}.

\section{RESULTS}
\subsection{Morphology}
\begin{figure*}[t]
\centering
\includegraphics[width=\textwidth]{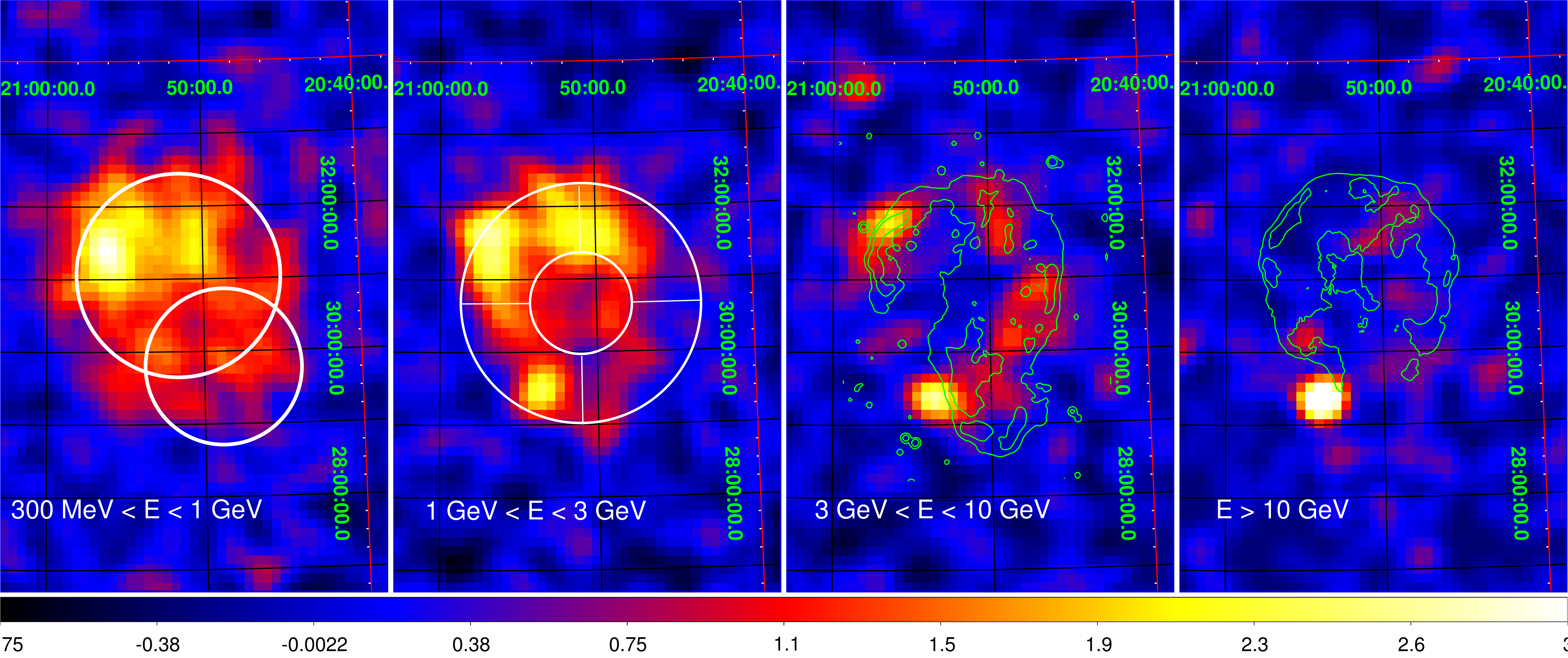} 
\caption{S/N maps of the null hypothesis in different energy ranges. From left to right, the panels include: the two regions discussed in section~\ref{sec:spec}; the ring used for modeling in \citet{cygloop}; the radio intensity contours from \citet{uyaniker}; and the X-ray contours from \citet{aschenbach}.}
\label{fig:CLmulti}
\end{figure*}

The S/N maps at different energy ranges, produced with the null hypothesis are shown in Figure~\ref{fig:CLmulti}. While an extended source is clearly seen at energies below 10\,GeV, in the last panel only residual, localized emission is present south of the SNR.

As a first step we reproduced the analysis from \citet{cygloop}. For this purpose we modeled the Cygnus Loop as a ring of 1.6/0.7 degree outer/inner diameter, centered at (RA, DEC) = ($20^{h}51^{m}$, $30\arcdeg50\arcm$). The ring is divided in four quadrants, and the spectral parameters of all of them are varied simultaneously. We note that the hard spot remaining above 10\,GeV is included in the southeast (SE) quadrant of the ring used in \citet{cygloop}. By substituting the SE quadrant by a point-like source, we find that LR improves by 124.
The position of this point-like source optimized by the tool \texttt{gtfindsrc} is (RA, DEC) = ($20^{h}53^{m}55^{s}$, $29\arcdeg24\arcm45\arcs$) with an uncertainty of $0.02\arcdeg$. We call this source J2053.9+2924. Its position is coincident with the X-ray and radio source 2E\,2051.7+2911, which is likely an AGN \citep{brink}. Therefore, we consider that J2053.9+2924 is a source in the background of the Cygnus Loop, and should not intervene in the modeling of the diffuse emission\footnote{The point-like source found in this analysis is called  3FGL\,J2053.9+2922 in the recently published Third \fermi\, Source Catalog \citep{3fgl}. We note that the source overlaps with the template for the Cygnus Loop, which in 3FGL is still modeled as the ring defined in \citet{cygloop}.}.

Having included J2053.9+2924 in the list of point-like sources, we maximize the likelihood of a template generated from the X-ray counts map observed by ROSAT \citep{aschenbach}, re-binned to match the pixel size of maps of the present analysis.
The likelihood ratios for the spatial models mentioned above are shown in Table~\ref{tab:CLtemplates}. It is clear that the thermal X-ray emission correlates very well with the observed gamma-ray emission, and requires less degrees of freedom than the ring to describe it.

\begin{table}[t]
\begin{center}
\caption{Likelihood ratio (LR) of the tested templates, with the number of degrees of freedom (d.o.f) added to the null hypothesis after selecting the best spectral model (Section~\ref{sec:spec}). Three of the additional d.o.f. always correspond to the spectral parameters of J2053.9+2924, except for (1), where original template by \citet{cygloop} is tested.}
\begin{tabular}{|l|c|c|c|}
\hline \textbf{Model} & \textbf{LR} & \textbf{d.o.f}
\\
\hline 0) Null hypothesis & 0 & 0 \\
\hline 1) Ring & 2069 & 12 \\
Divided in four quadrants & & \\
\hline 2) 3/4 Ring & 2193 & 12 \\
SE quadrant substituted by J2053.9+2924 & & \\
\hline 3) ROSAT template & 2204 & 6 \\
\hline 4) ROSAT template & 2238 & 8 \\
Divided in NE and SW & & \\
\hline
\end{tabular}
\label{tab:CLtemplates}
\end{center}
\end{table}

\begin{figure*} 
\includegraphics[width=0.9\textwidth, trim=1cm 1cm 0cm 0cm, clip=true]{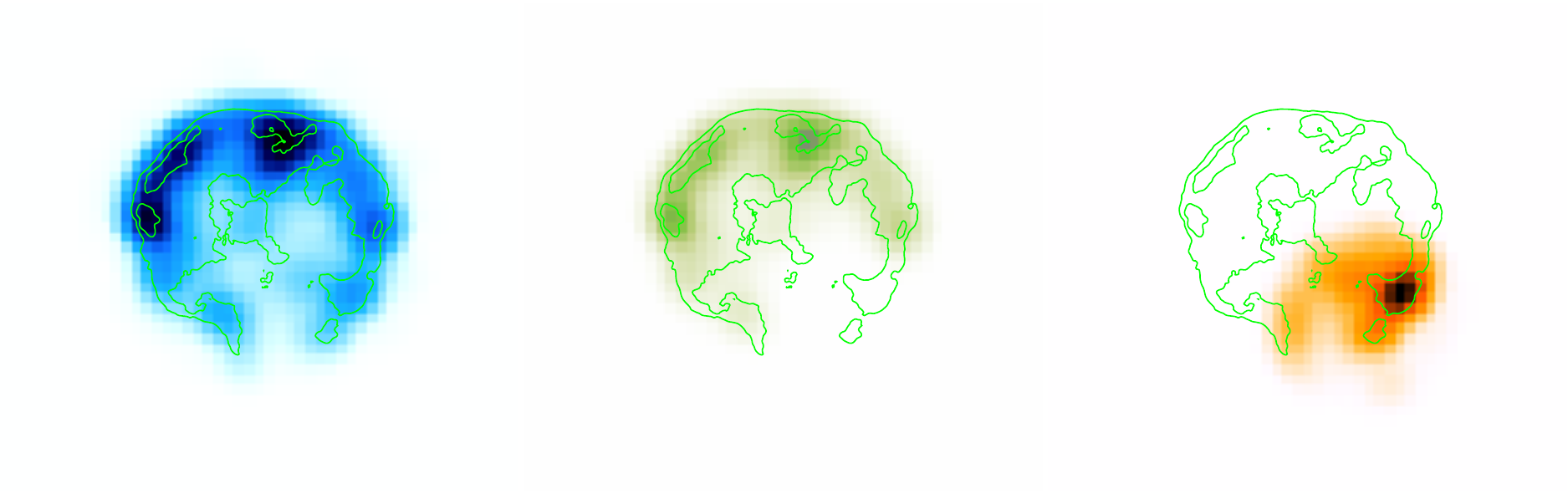} %
\caption{Sketch of the templates used in the \fermi\, analysis. From left to right: the total emission, the NE region, and the SW region. Color code matches that from Figure~\ref{fig:spec}. Green contours are the same X-ray contours as in Figure~\ref{fig:CLmulti} \citep{aschenbach}.}
\label{fig:templates}
\end{figure*}

The X-ray emission is very faint in the region of highly polarized radio emission, but we divide the spatial template in order to study this particular region. The templates for the main (NE) emission is cropped to avoid having pixels accounted for twice in the overlapping region (Figure~\ref{fig:templates}). We verify that both regions contribute significantly to the overall emission. Next, we proceed to test different the spectral models.

\subsection{Spectrum}
\label{sec:spec} 
\begin{figure} 
\includegraphics[trim=1.5cm 0.5cm 0cm 2cm, clip=true, height=0.5\textwidth, angle=-90]{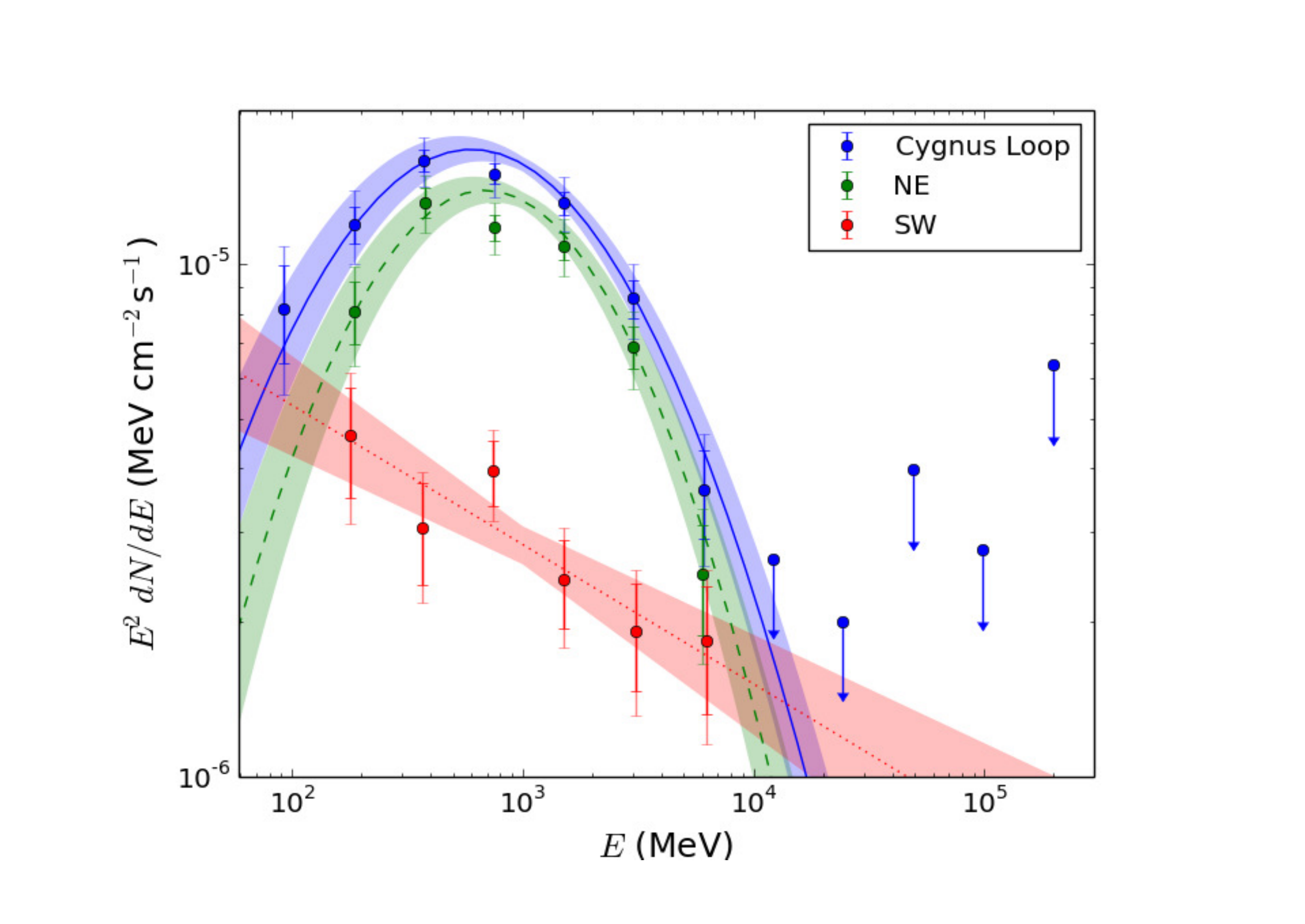}
\caption{Energy spectrum of the Cygnus Loop (blue). The emission from the NE region is shown in green, while the SW region is represented in red. The statistical uncertainty range of the best model is shown as a shaded area. Spectral points include statistical uncertainties (solid bars) and systematic uncertainties (shaded bars).}
\label{fig:spec}
\end{figure}

The energy spectrum of the source is shown in Figure~\ref{fig:spec}. The global emission, as well as the emission of the NE region are well described by a log-parabolic shape. To infer the spectral shape of the SW region, we test a power-law shape, a log-parabolic shape, and a power law with exponential cutoff. Because of the proximity of J2053+2923 to this region, we also test all three possible models for the point-like source. Then, we evaluate the likelihood ratio of each combination with respect to the initial assumption of both components having power-law shaped spectrum. We observe that models where J2053+2923 has an additional degree of freedom in the spectrum have a likelihood ratio with respect to the power-law/power-law hypothesis of $\sim10$, whereas models where the additional degree of freedom is added to the SW region only improve the likelihood ratio by $\sim1$. Therefore, we conclude that there is $\sim3\,\sigma$ evidence that the spectrum of J2053+2923 is curved, whereas the spectrum of the SW region is compatible with being a simple power law. The best fit spectral parameters of the global emission and the studied regions are shown in Table~\ref{tab:specpara}.

\begin{table}[t]
\begin{center}
\caption{Spectral parameters of the Cygnus Loop and its NE and SW regions under the assumption of a log-parabolic spectral shape, $dN/dE = N_0(E/1\,\mathrm{GeV})^{-\alpha-\beta\log(E/1\,\mathrm{GeV})}$. The photon flux at energies above 58.5\,MeV is shown in the last column.}
\begin{tabular}{|l|c|c|c|c|c|}
\hline \textbf{Region} & \textbf{$\alpha$} & \textbf{$\beta$} & \textbf{Flux} \\ 
& & & $10^{-8}$ph\,cm$^{-2}$s$^{-1}$ \\ 
\hline Cygnus Loop & $2.26\pm0.03$ & $0.25\pm0.03$ & $13.5\pm0.9$ \\ 
\hline NE          & $2.24\pm0.04$ & $0.32\pm0.04$ & $9.0\pm0.9$  \\ 
\hline SW          & $2.27\pm0.06$ & $0$           & $7.2\pm1.1$  \\ 
\hline
\end{tabular}
\label{tab:specpara}
\end{center}
\end{table}

\subsection{The point-like source J2053+2923}
\label{sec:j2053}
The point-like source South of the Cygnus Loop is detected with high significance (TS=214). As mentioned in Section~\ref{sec:spec}, the spectrum is described either by a power law with exponential cutoff or by a log-parabolic shape, but the power law with exponential cutoff hypothesis is slightly preferred. The cutoff energy is $(22\pm10_{stat})$\,GeV, while the spectral index below the cutoff is $1.46\pm0.18_{stat}$. The source is not significantly detected at low energies. Using the same spectral binning as for the Cygnus Loop, the flux is measurable (TS$_{bin}>10$) between 1\,GeV and 72\,GeV, making this source a candidate very-high-energy emitter. We also performed an unbinned likelihood analysis in time intervals of 60 days. The source is detected with TS$_{2month}>10$ in 13 out of 37 such intervals. This hint of variability supports the association with the background AGN, 2E\,2051.7+2911.

\section{CONCLUSIONS}
Due to its proximity and angular size, the Cygnus Loop permits spatially resolved studies of different parts of the SNR, that interact with different components of surrounding medium. Particularly, it is known that the NE of the shell interacts with relatively dense medium and is thus bright in X-rays and optical emission lines compared to other parts of the shell. These inhomogeneities are likely to happen in other remnants from core-collapse supernovae, while remaining unnoticed due to lack of resolution of the instruments.
Understanding the physical mechanisms that power the gamma-ray emission of the Cygnus Loop, and the differences between different regions of the shell, may help understand the variety of spectral shapes that SNRs display at gamma-ray energies.

The fact that most of the gamma-ray emission from the Cygnus Loop follows closely the thermal X-ray emission from shocked matter supports the idea that most of its gamma-ray emission is emitted by interactions of accelerated hadrons with the dense medium. In this case, a low-energy break is expected in the spectrum due to the production threshold of neutral pions \citep{w44ic443}.This is the case for the energy spectrum measured in this analysis, which has a maximum around 0.6\,GeV. However, the SW portion of shell (which is brighter in radio and fainter in X-rays), has a different gamma-ray spectrum without indication of a spectral break. The explanation for the different gamma-ray properties, and the related radiative processes of the two regions, including the two-SNR scenario, is under investigation.

\bigskip 

\begin{acknowledgments}
This work was supported by the interdisciplinary project \texttt{Labex UnivEarthS} from Sorbonne Paris cit\'e (ANR-10-LABX-0023 and ANR- 11-IDEX-0005-02). We also acknowledge support by NSERC through the Canada Research Chairs and Discovery Grants programs.
\end{acknowledgments}

\bigskip 

\bibliography{bibliography}


\end{document}